\documentclass[prb,twocolumn,showpacs,floatfix,superscriptaddress]{revtex4-1}

\usepackage{graphicx}
\usepackage{dcolumn}
\usepackage{amsmath}
\usepackage{color}

\begin{document}
\title{Electron spin resonance and exchange paths in the orthorhombic dimer system Sr$_2$VO$_4$}

\author{J.~Deisenhofer}
\author {S.~Schaile}
\affiliation{Experimentalphysik V, Center for Electronic
Correlations and Magnetism, Institute for Physics, Augsburg
University, D-86135 Augsburg, Germany}
\author{J.~Teyssier}
\affiliation{D\'{e}partement de Physique de la Mati\`{e}re
Condense´e, Universit\'{e} de Gen\`{e}ve, CH-1211 Gen\`{e}ve 4,
Switzerland}
\author{Zhe Wang}
\author{M.~Hemmida}
\author{H.-A.~Krug von Nidda}
\affiliation{Experimentalphysik V, Center for Electronic
Correlations and Magnetism, Institute for Physics, Augsburg
University, D-86135 Augsburg, Germany}
\author{R.~M.~Eremina}
\affiliation{E. K. Zavoisky Physical Technical Institute, 420029
Kazan, Russia}
\author{M.~V.~Eremin}
\affiliation{Institute for Physics, Kazan (Volga region) Federal
University, 430008 Kazan, Russia}
\author{R.~Viennois}
\affiliation{D\'{e}partement de Physique de la Mati\`{e}re
Condense´e, Universit\'{e} de Gen\`{e}ve, CH-1211 Gen\`{e}ve 4,
Switzerland}
\author{E.~Giannini}
\affiliation{D\'{e}partement de Physique de la Mati\`{e}re
Condense´e, Universit\'{e} de Gen\`{e}ve, CH-1211 Gen\`{e}ve 4,
Switzerland}
\author{D.~van der Marel}
\affiliation{D\'{e}partement de Physique de la Mati\`{e}re
Condens\'{e}e, Universit\'{e} de Gen\`{e}ve, CH-1211 Gen\`{e}ve 4,
Switzerland}
\author{A.~Loidl}
\affiliation{Experimentalphysik V, Center for Electronic
Correlations and Magnetism, Institute for Physics, Augsburg
University, D-86135 Augsburg, Germany}

\date{\today}

\begin{abstract}

We report on magnetization and electron spin
resonance (ESR) measurements of Sr$_{2}$VO$_4$ with orthorhombic symmetry. In this
 dimer system  the $V^{4+}$ ions are in tetrahedral
environment and are coupled by an antiferromagnetic intra-dimer
exchange constant $J/k_B \approx$ 100~K to form a singlet ground state
without any phase transitions between room temperature and 2~K. Based on an extended-H\"{u}ckel-Tight-Binding analysis we identify the strongest exchange interaction to occur between two inequivalent vanadium sites via two intermediate oxygen ions.
The ESR absorption spectra can be well
described by a single Lorentzian line with an effective g-factor $g$
= 1.89. The temperature dependence of the ESR intensity is well
described by a dimer model in agreement with the magnetization data. The temperature dependence of the ESR linewidth can be modeled by a superposition of a linear increase with temperature with a slope $\alpha$ = 1.35 Oe/K and a thermally
activated behavior with an activation energy $\Delta/k_B$ = 1418~K, both of which point to spin-phonon coupling as the dominant relaxation mechanism in this compound.

\end{abstract}



\maketitle

\section{Introduction}

Quantum magnetism is a fascinating research field with a plethora of
observed and predicted exotic phenomena such as the Bose-Einstein
condensation of magnons\cite{Giamarchi08} or quantum
spin-liquids.\cite{Balents10} In transition-metal oxides where the
magnetic ions are in 3$d^1$ or 3$d^9$ electronic configuration with
spin $S$ = 1/2 such as, for example, Cu$^{2+}$ cuprates, Ti$^{3+}$
in titanates or V$^{4+}$ in vanadates, the coupling of spin,
orbital, and lattice degrees of freedom makes the ground-state
properties particularly rich and complex.\cite{Tokura00,Zhou07,Jackeli09,Viennois10,Zhou10,Teyssier11,Eremin11}

\begin{figure}[t]
\includegraphics[width=0.45\textwidth,clip]{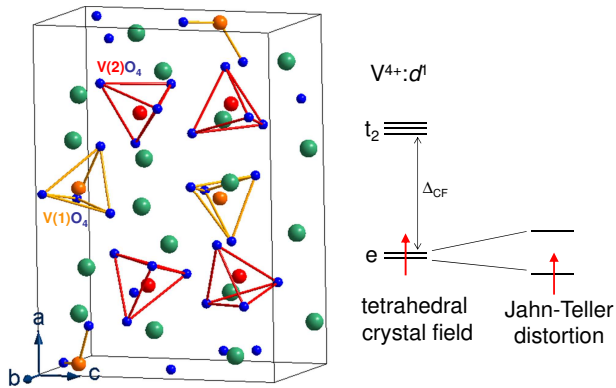}
\caption{\label{fig:structureortho} Left: Unit cell of orthorhombic Sr$_2$VO$_4$ with space group \textit{Pna2$_1$} (Ref.~\onlinecite{Gong91}), showing the tetrahedral coordination of the two inequivalent vanadium sites V(1) and V(2).
Right: Schematic of the splitting of the V$^{4+}$ $d$-levels as described in the text.}
\end{figure}

In this study we will focus on orthorhombic Sr$_2$VO$_4$, where the V$^{4+}$ ions are in
3$d^1$ configuration and the electron occupies the low-lying
$e$-states in
tetrahedral environment as sketched in Fig.~\ref{fig:structureortho}.  The material exhibits orthorhombic symmetry with space group \textit{Pna2$_1$} and lattice parameters \textit{a} =
14.092(4) {\AA}, \textit{b} = 5.806(2) {\AA}, and \textit{c} = 10.106(3) {\AA} (see Fig.~\ref{fig:structureortho}).\cite{Gong91} The orthorhombic distortion can be interpreted in terms of a Jahn-Teller distortion which removes the orbital degeneracy of the V$^{4+}$ ions. No phase transitions have been observed in the
temperature range from 4 to 300~K for orthorhombic Sr$_2$VO$_4$. Its susceptibility has been
described in terms of a spin-dimer system with a singlet ground state
and an antiferromangetic intra-dimer coupling of about 100~K.\cite{Gong91} However, a
clear identification of the superexchange paths corresponding to the
magnetic intra-dimer coupling is not available at present, because
the superexchange paths between the structural VO$_4$ units will
involve two or more ligands. Such more complicated
super-superexchange (SSE) paths have been found to yield exchange
couplings of considerable magnitude and to determine the ground
state properties in a large number of
compounds.\cite{Whangbo03,Deisenhofer06,Eremina11,Wang11JPSJ}

Here we investigate orthorhombic Sr$_2$VO$_4$ by magnetization and electron spin resonance experiments. The exchange paths are analysed by an extended-H\"{u}ckel-Tight-Binding (EHTB) approach and one dominant exchange path via two intermediate oxygen ions is identified. The ESR intensity confirms the dimer-picture for the susceptibility, the spin-orbit coupling is estimated from the effective $g$-factor, and the linewidth seems to be governed by a phonon-mediated relaxation mechanism and a thermally activated process.

\section{Experimental Details}

Ceramic samples were prepared from a Sr$_4$V$_2$O$_9$ precursor\cite{Viennois10} by four consecutive reduction and grinding processes at 1100 °C in sealed quartz tubes with metallic Zr as an oxygen getter. The samples were characterized by X-ray powder diffraction and showed good agreement with the reported symmetry and lattice parameters.\cite{Gong91} Susceptibility measurements were
performed using a SQUID magnetometer (Quantum Design).
ESR measurements were performed in a Bruker ELEXSYS E500 CW-spectrometer
at X-band frequencies ($\nu \approx$ 9.47 GHz) equipped with a
continuous He-gas-flow cryostat in the temperature region $4<T< 300$
K. ESR detects the power $P$ absorbed by the sample from the
transverse magnetic microwave field as a function of the static
magnetic field $H$. The signal-to-noise ratio of the spectra is
improved by recording the derivative $dP/dH$ using lock-in technique
with field modulation.

\section{Experimental Results and Discussion}

\subsection{Magnetic Susceptibility}

\begin{figure}[b]
\includegraphics[width=0.45\textwidth,clip]{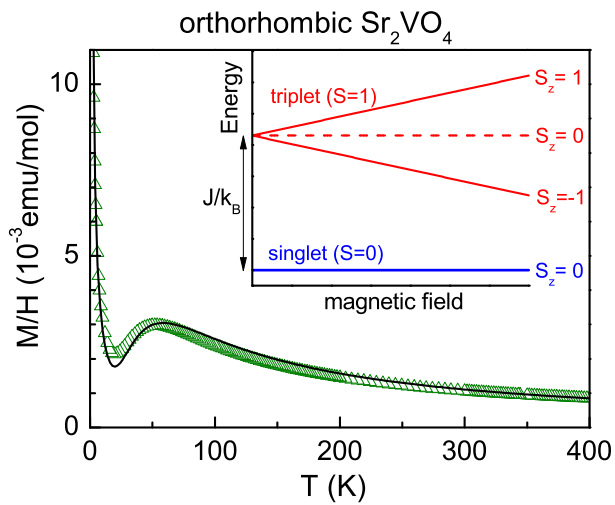}
\caption{\label{fig:chi_ortho} Temperature dependence of the magnetic susceptibility $M/H$ for orthorhombic Sr$_2$VO$_4$ measured in a magnetic field of $\mu_0H$ = 0.1\;T. The solid line is a fit using Eq.~(\ref{eq:chi}). The inset shows the corresponding energy level scheme of a spin dimer with antiferromagnetic exchange coupling $J$ as a function of the applied field $H$.}
\end{figure}

Let us now consider the susceptibility of Sr$_2$VO$_4$ as shown in Fig.~\ref{fig:chi_ortho}. This system has been described
previously as a system of antiferromagnetically coupled spin-dimers
with a singlet ground state.\cite{Gong91} To analyze the susceptibility $\chi=M/H$ determined from the magnetization $M$ divided by the applied magnetic field $H$ in the entire temperature range we use
\begin{equation}\label{eq:chi}
 \chi=\chi_{0} + \chi_C + \chi_{BB},
\end{equation}
with a Curie contribution $\chi_C=C/T$ due to unbound spins and
magnetic impurities and a constant contribution $\chi_0$, and the dimer susceptibility $\chi_{BB}$ as
derived by Bleaney and Bowers:\cite{Bleaney52}
\begin{equation}\label{eq:BB}
\chi_{BB}(T)=\frac{Ng^2\mu_B^2}{k_B T}[3+ \exp{(J/k_BT)}]^{-1}.
\end{equation}
Here $J$ denotes the intradimer exchange coupling, $g$ is the
effective $g$-factor of the vanadium ions, and $\mu_B$ is the Bohr
magneton. The $g$-factor was fixed to the experimental value
$g=1.89$ observed in the ESR measurements (see below). The obtained
fit parameters are $J=104~K$, $\chi_0$=-1$\cdot10^{-4}$ emu/mol, which is of the typical order of magnitude for a diamagnetic contribution, and
$C$=0.028 emuK/mol, corresponding to about 7\% of unpaired spins. The value for the intradimer exchange $J=104~K$
is in agreement with literature\cite{Gong91} and corresponds nicely
to a magnetic excitation peaked at 8.6~meV observed by neutron
scattering.\cite{Toth12} From the structural arrangement, however, it is not clear which of the possible exchange paths corresponds to this dominant exchange-coupling constant. Therefore, we performed an extended H\"{u}ckel-Tight-Binding analysis of the exchange paths which will be discussed in the following.


\subsection{Analysis of the exchange paths}

\begin{figure}[t]
\centering
\includegraphics[width=80mm,clip]{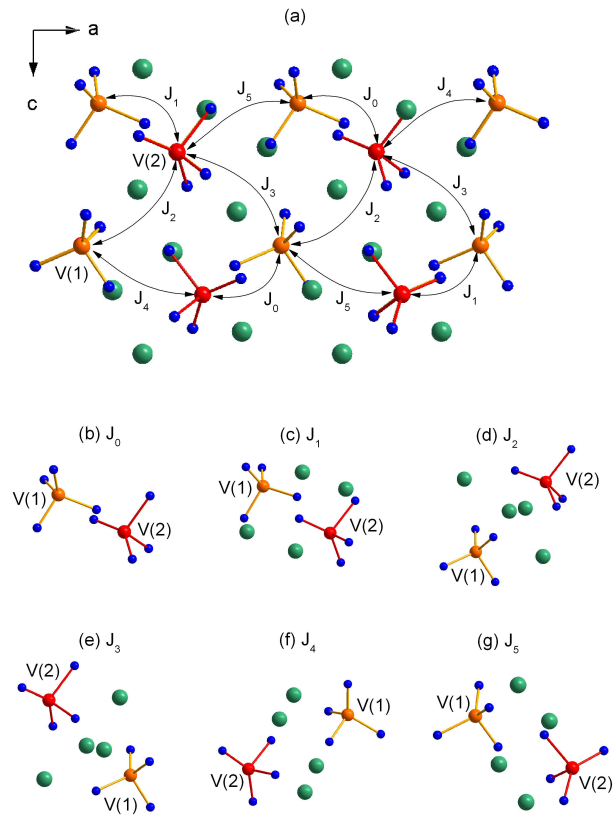}
\vspace{2mm} \caption[]{\label{Fig:ExchangeStruc} (Color online) (a) Projection of
orthorhombic lattice structure of Sr$_2$VO$_4$~with space group $Pna2_1$ on the \emph{ac}-plane.\cite{Gong91}
The exchange paths between neighboring V ions are denoted by the corresponding exchange constants $J_0$--$J_5$ in the sequence of increasing V$\cdots$V distance.
(b)--(g) Spin dimers associated with these exchange paths.  The large, middle, and small spheres show Sr, V, and O atoms, respectively.
The two crystallographically inequivalent V ions are denoted as V(1) and V(2), respectively.}
\end{figure}

Six distinct exchange paths with exchange couplings $J_0$--$J_5$ and increasing distance between the vanadium ions can be identified in the structure of Sr$_2$VO$_4$ (see Fig.~\ref{Fig:ExchangeStruc} and Table~\ref{Tab:Js}).

The interaction between the magnetic orbitals of two ions in a spin
dimer gives rise to two molecular orbitals with an energy split
$\Delta e$. In the spin-dimer analysis based on EHTB
calculations,\cite{Whangbo03, SAMOA} the strength of an
antiferromagnetic exchange interaction between two spin sites is
estimated by $J_{AF}=-(\Delta e)^2/U_{eff}$, where $U_{eff}$ is the
effective on-site repulsion that is nearly constant for a given
compound.

Double-$\zeta$ Slater-type orbitals are adopted to describe the
atomic \emph{s}, \emph{p}, and \emph{d} orbitals in the EHTB
calculations.\cite{Whangbo03} The atomic parameters used for the
present EHTB calculations of $(\Delta e)^2$ are summarized in
Table~\ref{Tab:parameter}. The parameters of V and O atoms are
referred to the previous EHTB calculations on other vanadate
compounds,\cite{Koo02} while the rest are taken from the atomic
orbital calculations.\cite{Clementi74, Whangbo03}

\begin{table}[h]
\caption{\label{Tab:parameter} Exponents $\zeta_i$ and valence shell
ionization potentials $H_{ii}$ of Slater-type orbitals $\phi_i$ used
for extended H\"{u}ckel tight-binding calculations. $H_{ii}$ are the
diagonal matrix elements $\langle \phi_i |H_{eff}| \phi_i \rangle$,
where $H_{eff}$ is the effective Hamiltonian. For the calculation of
the off-diagonal matrix elements $H_{ij}=\langle \phi_i |H_{eff}|
\phi_j \rangle$, the weighted formula as described in
Ref.~\onlinecite{Ammeter78} was used. $C$ and $C'$ denote the
contraction and diffuse coefficients used in the double-$\zeta$
Slater-type orbitals.\cite{Whangbo03, Koo02, Clementi74}}
\begin{ruledtabular}\vspace{0.5mm}
\begin{tabular}{c@{\hspace{3em}}*{7}{c}}
atom    &$\phi_i$    &$H_{ii}$    &$\zeta_i$   &$C$       &$\zeta_{i'}$      & $C'$
\vspace{0.5mm}\\\hline
V       &4s          &-8.81       &1.697       &1.0
\\
V       &4p          &-5.52       &1.260       &1.0
\\
V       &3d          &-11.0       &5.052       &0.3738     &2.173            &0.7456
\\
Sr      &5s          &-6.62       &1.630       &1.0
\\
Sr      &5p          &-3.92       &1.214       &1.0
\\
O       &2s          &-32.3       &2.688       &0.7076     &1.675            &0.3745
\\
O       &2p          &-14.8       &3.694       &0.3322     &1.825            &0.7448
\\
\end{tabular}
\end{ruledtabular}
\end{table}

As suggested in Ref.~\onlinecite{Gong91}, the exchange paths between
neighboring vanadium atoms could be V--O--O--V or V--O--Sr--O--V.
According to our calculations, the interactions between second- to
sixth-nearest neighboring pairs of V ions are significantly
increased, when the strontium atoms are considered in the exchange
paths. Therefore, the exchange paths are chosen as V--O--Sr--O--V for
$J_1$--$J_5$ (see Fig.~\ref{Fig:ExchangeStruc} (c)-(g)). In
contrast, strontium atoms are not considered for the exchange path
$J_0$.

\begin{table}[h]
\caption{\label{Tab:Js} Values of the V$\cdots$V distance in {\AA} and
$(\Delta e)^2$ associated with the exchange path $J_0$--$J_5$ in
Sr$_2$VO$_4$}
\begin{ruledtabular}\vspace{0.5mm}
\begin{tabular}{c@{\hspace{3em}}*{4}{c}}
path &\multicolumn{1}{c}{V$\cdots$V}  &\multicolumn{1}{c}{$(\Delta e)^2$}  &$J_i/J_0$
\vspace{0.5mm}\\\hline
$J_0$ &\multicolumn{1}{c}{4.090}  &\multicolumn{1}{c}{1370}  &1.00
\\
$J_1$  &4.682   &46       &0.03
\\
$J_2$  &4.734   &210      &0.15
\\
$J_3$  &4.978   &55       &0.04
\\
$J_4$  &5.381   &69       &0.05
\\
$J_5$  &5.511   &15       &0.01
\\

\end{tabular}
\end{ruledtabular}
\end{table}

The results of our calculations are summarized in
Table~\ref{Tab:Js}. It shows that the dominant spin-dimer exchange
is mediated by the path corresponding to $J_0$ (see Fig.~\ref{Fig:ExchangeStruc}(b)),
where the distance between two V ions is shortest. It is
interesting that the second strongest exchange is not between the
second-nearest-neighbor V ions, but mediated along the path with $J_2=0.15J_0$.

\subsection{Electron Spin Resonance }\label{ESRTheory}

\begin{figure}[tbh]
\includegraphics[width=0.45\textwidth]{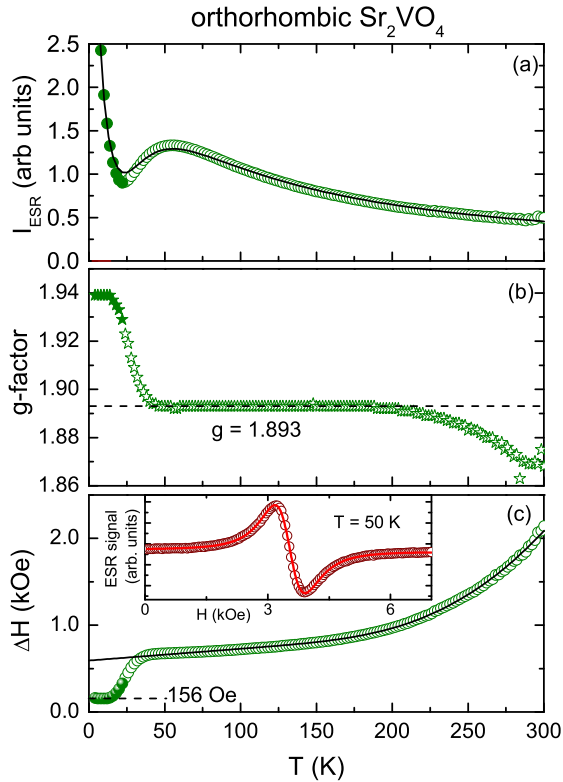}
\caption{\label{fig:ESRdataortho} Temperature dependence of (a) the ESR
spin susceptibility together with a fit using Eq.~(\ref{eq:chi}), (b) the
effective $g$-factor, and (c) the ESR linewidth in Sr$_2$VO$_4$ together with a fit using Eq.~(\ref{eq:deltaH}).}
\end{figure}

The absorption spectra of Sr$_2$VO$_4$ can be described by an
exchange-narrowed Lorentzian line shape as shown in the inset of
Fig.~\ref{fig:ESRdataortho}(c). The temperature dependences of the
obtained fit parameters are shown in Fig.~\ref{fig:ESRdataortho}.

The temperature dependence of the ESR intensity $I_{ESR}$ can be well fitted by using Eq.~(\ref{eq:chi}) (solid line in Fig.~\ref{fig:ESRdataortho}(a)) and yields a slightly larger exchange constant $J$ = 107~K. The experimental effective $g$-factor of 1.89 as  shown in Fig.~\ref{fig:ESRdataortho}(b) was used and $\chi_{0}$ was set to zero, because it does not contribute to the resonance absorption. These parameters are in agreement with the fit for $M/H$ and show that the resonance absorption originates from magnetic-dipole allowed intra-triplet excitations with $\Delta S_z= \pm 1$ (see inset of Fig.~\ref{fig:chi_ortho}). The increase of the $g$-factor and the decrease of the ESR linewidth below 30~K signal the depopulation of the excited triplet state and the ESR intensity should drop to zero in the ground state. Instead, the intensity increases towards lower temperatures in a Curie-like fashion (solid points) indicating that the resonance signals at lowest temperatures with $g=1.94$ and $\Delta H =156$~Oe belong to unpaired paramagnetic ions in the sample.

For the intra-triplet excitations (open symbols) we find an almost temperature independent $g$-factor $g$ = 1.89
between 50 and 150~K. The decrease towards higher temperatures is
probably related to the increasing linewidth, which reaches the order
of magnitude of the resonance field above 200~K and, hence, imposes
a larger uncertainty on the resonance field (or $g$-factor) as a
fitting parameter. In first-order perturbation theory the effective
$g$-factor is given by
\begin{equation} \label{gfac-ortho}
g = 2- \frac{4\lambda}{\Delta_{CF}},
\end{equation}
with the spin-orbit coupling $\lambda$ and the $e-t_2$ crystal-field
splitting parameter $\Delta_{CF}$.\cite{Abragam70} Using $g$ = 1.89 and
$\Delta_{CF}$ = 8900 cm$^{-1}$ as observed by ellipsometry
measurements\cite{Teyssier12} we estimate $\lambda$ = 244 cm$^{-1}$
(30~meV) in good agreement with the value obtained for V$^{4+}$ ions.\cite{Abragam70,Jackeli09,Eremin11}

The temperature dependence of the ESR linewidth $\Delta H$ of the intra-triplet excitations is shown
in Fig.~\ref{fig:ESRdataortho}(c). The linewidth increases monotonously with temperature, between 50~K and 170~K only with a moderate slope but for higher temperatures a strong increase sets in, indicating the presence of at further relaxation mechanisms.
The temperature dependence can be well described by
\begin{equation} \label{eq:deltaH}
\Delta H = \Delta H_0 +\alpha T + Ae^{-\frac{\Delta}{k_BT}},
\end{equation}
with $\Delta$ = 1418(19)~K, a residual zero-temperature value $\Delta H_0$ = 593(5)~Oe, $\alpha$ = 1.35(4) Oe/K, and $A$ = 1.23(7)$\cdot 10^5$ Oe.

The linear term can be understood in terms of a spin-phonon relaxation mechanism, where one phonon is involved in the relaxation process.\cite{Abragam70,Seehra68,Castner71} The relaxation rate depending on the probabilities for absorption and emission of the phonon will follow a $\coth(\hbar\omega_{ph}/k_BT)$-behavior which yields a linear behavior for $\hbar\omega_{ph}/k_BT\ll 1$. Hence, any possible source of line-broadening such as a Dzyaloshinsky-Moriya (DM) or symmetric anisotropic exchange interaction, which might be directly modulated by one phonon, could be the origin of the linear contribution.\cite{Abragam70,Seehra68,Castner71,Zorko04} We want to mention that we have no indication of the presence of a sizeable static DM interaction. However, there is no center of inversion between the two inequivalent V sites and a static DM contribution within the dimers could arise.


A thermally activated contribution has been observed for several low-dimensional magnets \cite{Heinrich03,Zakharov06,Eremin08}  and in the dimer system Sr$_3$Cr$_2$O$_8$, where the Cr$^{5+}$ also have an electronic 3$d^1$ configuration in a tetrahedral crystal field.\cite{Wang11} In the latter compound the value of $\Delta$ = 388~K is lying within the phonon frequency range and the contribution was tentatively assigned to stem from a two-phonon Orbach process, where the spin relaxation occurs via an absorption of a phonon to a higher-lying electronic state in the energy range of the phonon continuum. For Sr$_2$VO$_4$ the value $\Delta$ = 1418(19)~K is too high for phonon modes and rules out the presence of an Orbach mechanism. Since all of the mentioned studies deal with Jahn-Teller active ions, another possible origin of such a thermally activated behavior could be the presence of different Jahn-Teller distortions, which are close in energy, e.g. in case of the one-dimensional magnet CuSb$_2$O$_6$ (Cu$^{2+}$ with spin 1/2 in octahedral environment) the exponential increase of the linewidth with $\Delta$ = 1484~K has been observed on approaching a static-to-dynamic Jahn-Teller transition at 400\,K.\cite{Heinrich03} The value of $\Delta$ would then correspond to the energy barrier separating the two Jahn-Teller configurations.

%
%
%
%

%

\section{Summary}

In summary, we investigated orthorhombic Sr$_{2}$VO$_4$ by electron spin
resonance measurements and identified the dominating exchange path to occur between two inequivalent vanadium sites via two intermediate oxygen ions using an extended-H\"{u}ckel-tight binding analysis. The temperature dependence of the ESR intensity and the magnetization reveal a dimerized singlet ground state with an intradimer coupling constant $J/k_B \approx$ 100~K.
The ESR linewidth exhibits an increase with rising temperature which can be understood in terms of a phonon-modulated spin relaxation yielding a linear increase with slope $\alpha$ = 1.35 Oe/K and a thermally
activated Arrhenius behavior with an activation energy $\Delta/k_B$ = 1418~K, which might be related to the Jahn-Teller distortion of the system.

\begin{acknowledgments}
We thank D. Vieweg for experimental support and H.-J. Koo and M.-H.
Whangbo for fruitful discussions with regard to the EHTB
calculation. This work is partially supported by the SNSF through
Grant No. 200020-130052 and the National Center of Competence in
Research (NCCR) "Materials with Novel Electronic Properties-MaNEP"
and by the DFG via the Collaborative Research Center TRR 80
(Augsburg-Munich) and project DE 1762/2-1.
\end{acknowledgments}


\end{document}